\documentclass[conference]{IEEEtran}
\usepackage{amsmath,graphicx}

\usepackage{subfig}
\usepackage{algorithm,algorithmic}
\usepackage{enumerate}
\usepackage{amssymb}
\usepackage{amsthm}
\floatname{algorithm}{The $q$-SF Algorithm}

\newtheorem{theorem}{Theorem}[section]
\newtheorem{lemma}[theorem]{Lemma}
\newtheorem{proposition}[theorem]{Proposition}
\newtheorem{corollary}[theorem]{Corollary}
\newtheorem{assumption}{Assumption}

\title{\lowercase{$q$}-Gaussian based Smoothed Functional \\ Algorithms for Stochastic Optimization}
\author{\IEEEauthorblockN{Debarghya Ghoshdastidar, Ambedkar Dukkipati and Shalabh Bhatnagar}
\IEEEauthorblockA{Department of Computer Science and Automation\\
Indian Institute of Science, Bangalore -- 560012 \\
Email: \{gdebarghya@ee,ambedkar@csa,shalabh@csa\}.iisc.ernet.in}}

\begin{document}
\maketitle

%======================Abstract=================
\begin{abstract}
The $q$-Gaussian distribution results from maximizing certain
generalizations of Shannon entropy under some constraints. The importance of $q$-Gaussian
distributions stems from the
fact that they exhibit power-law behavior, and also generalize Gaussian
distributions. In this paper,
we propose a Smoothed Functional (SF) scheme for gradient estimation using $q$-Gaussian distribution,
and also propose an algorithm for optimization based on the above scheme. Convergence results of
the algorithm are presented. Performance of the proposed algorithm is 
shown by simulation results on a queuing model.
\end{abstract}

%====================Section: Introduction================
\section{Introduction}

Stochastic optimization algorithms play an important role in optimization problems involving objective functions that cannot 
be computed analytically. These schemes are extensively used in discrete event systems, such as
queuing systems, for obtaining optimal or near-optimal performance measures.

Gradient descent algorithms are used for stochastic optimization by estimating the gradient of average cost in
the long run.
Methods for gradient estimation by random perturbation of parameters have been proposed in \cite{kiefer}.
The Smoothed Functional (SF) scheme, 
described in \cite{katkovnik}, approximates the gradient of expected cost by its convolution with a multivariate normal
distribution. Based on all the above schemes, two-timescale stochastic approximation algorithms have been presented in \cite{bhatnagar3},
which simultaneously perform cost averaging and parameter updation using different step-size schedules.
The main issue with such algorithms is that, although convergence 
to a local optimum is guaranteed, the global optimum cannot achieved in practice.
Hence, new methods are sought. 

In this paper, we propose a new SF technique based on $q$-Gaussian distribution, which is a generalization of the Gaussian distribution.
We show that $q$-Gaussian satisfies all the conditions for smoothing kernels proposed by Rubinstein~\cite{rubinstein}. We illustrate  
a method for gradient estimation using $q$-Gaussian. We also present a two-timescale algorithm for stochastic optimization
using $q$-Gaussian based SF, and show the convergence of the proposed algorithm.

The rest of the paper is organized as follows. The framework for the optimization problem
and some of the preliminaries are presented in Section~\ref{background}. 
Gradient estimation using $q$-Gaussian SF has been derived in Section~\ref{validity}. Section~\ref{algorithm} presents the proposed
algorithm. Numerical experiments comparing our algorithm with a previous algorithm
is presented in Section~\ref{results}. An outline of convergence analysis of our algorithm is discussed in Section~\ref{convergence}.
Finally, Section~\ref{conclusion} provides the concluding remarks.   

%====================Section: Background and Preliminaries===================
\section{Background and Preliminaries}
\label{background}
\subsection{$q$-Gaussian distribution}
Most of the distributions, like normal, uniform, exponential etc., can be obtained by maximizing Shannon entropy functional defined as
{$H(p) = \int\limits_{\mathcal{X}}p(x)\mathrm{ln}p(x)\mathrm{d}x$},
where {$p$} is a pdf defined on the sample space {$\mathcal{X}$}. 
Other entropy functions have also been proposed as generalized
information measures. One of the most popular among them is nonextensive entropy, 
first introduced in~\cite{havrda}, and later studied by Tsallis~\cite{tsallis88}. 
Its continuous form entropy functional,
which is consistent with the discrete case~\cite{dukkipati7P}, is defined as
\begin{equation}
H_{q}(p) = \frac{1-\displaystyle\int\limits_{\mathcal{X}}[p(x)]^q\mathrm{d}x}{q-1}, \qquad{q\in\mathbb{R}}.
\end{equation}
This entropy functional produces Shannon entropy as {$q\rightarrow1$}.
% and also satisfies the generalized Shannon-Khinchin axioms in the discrete case~\cite{suyari}.
Corresponding to this generalized measure,
$q$-expectation of a function $f(.)$ can be defined as
\begin{equation}
\label{q-expect-defn}
\langle{f(x)}\rangle_q = 
\frac{\displaystyle\int\limits_{\mathbb{R}}f(x)[p(x)]^q\mathrm{d}x}
{\displaystyle\int\limits_{\mathbb{R}}[p(x)]^q\mathrm{d}x}.
\end{equation}
Maximizing Tsallis entropy under the following constraints:
\begin{equation}
\label{Gq1:constraints}
\langle{x}\rangle_q = \mu\qquad and \qquad \langle{x^2}\rangle_q = \beta^2,
\end{equation}
results in $q$-Gaussian distribution \cite{prato}, which is of the form
\begin{equation}
\label{Gq1:formula}
G_{q,\beta}(x) = {\frac{1}{{\beta}K_{q}}}\left(1-{\frac{(1-q)}{(3-q)\beta^2}(x-\mu)^2}\right)_+^{\frac{1}{1-q}},
\end{equation}
where {$y_+=\max(y,0)$} is called Tsallis cut-off condition,  
and {$K_q$} is the normalizing constant, which depends on the value of $q$. 
The function defined in~\eqref{Gq1:formula} is not integrable for {$q\geqslant3$}, and 
hence, $q$-Gaussian is a probability density function only for {$q<3$}. 
Multivariate form of the $q$-Gaussian distribution \cite{vignat} is defined as
\begin{equation}
\label{Gq:formula}
G_{q,\beta}(X) = {\frac{1}{\beta^NK_{q,N}}}\left(1-{\frac{(1-q)}{(3-q)\beta^2}{\Vert}X{\Vert}^2}
\right)_+^{\frac{1}{1-q}},
\end{equation}
where {$K_{q,N}$} is the normalizing constant. It is easy to verify that
the multivariate normal distribution is a 
special case of~\eqref{Gq:formula} as {$q\to1$}.
A similar distribution can also be obtained by maximizing R\'{e}nyi entropy~\cite{costa}.

\subsection{Problem Framework}
Let {$\{Y_n\}_{n\in\mathbb{N}} \subset\mathbb{R}^d$} be a parameterized Markov process, depending on a tunable
parameter {$\theta\in C$}, where {$C$} is a compact and convex subset of {$\mathbb{R}^N$}. Let {$\mathbb{P}_{\theta}(x,\mathrm{d}y)$} 
denote the transition kernel of {$\{Y_n\}$} when the operative parameter is {$\theta\in C$}. 
Let {$h:\mathbb{R}^d\mapsto\mathbb{R}^+\bigcup \{0\}$} be a Lipschitz continuous cost function associated with the process.
\begin{assumption}
\label{ergodic}
The process {$\{Y_n\}$} is ergodic for any given {$\theta$} as the operative parameter, \textit{i.e.},
\begin{displaymath}
\frac{1}{L}\sum_{m=0}^{L-1}h(Y_m) \to \mathbb{E}_{\nu_{\theta}}[h(Y)] \text{ as } L\to\infty,
\end{displaymath}
where {$\nu_{\theta}$} is the stationary distribution of {$\{Y_n\}$}.
\end{assumption}
Our objective is to minimize the long-run average cost
\begin{equation}
\label{Jdefn}
J(\theta) = \lim_{L\to\infty}\frac{1}{L}\sum_{m=0}^{L-1}h(Y_m) = \int\limits_{\mathbb{R}^d}h(x)\nu_{\theta}(\mathrm{d}x)
\end{equation}
by choosing an appropriate {$\theta\in C$}. The existence of the above limit is given by Assumption~\ref{ergodic}. 
In addition, we assume that the average cost {$J({\theta})$} satisfies the following condition.
\begin{assumption}
\label{differentiable}
{$J(\theta)$} is continuously differentiable with respect to any {$\theta\in{C}$}.
\end{assumption}
We also assume the existence of a stochastic Lyapunov function through the following assumption.
\begin{assumption}
\label{lyapunov}
Let {$\{\theta(n)\}$} be a sequence of random parameters, obtained using an iterative scheme, controlling the process {$\{Y_n\}$},
and {$\mathcal{F}_n = \sigma(\theta(m),$} {$Y_m,m\leqslant n)$}, {$n\geqslant0$} denote the sequence of associated {$\sigma$}-fields. 
\\There exists {$\epsilon_0>0$}, {$\mathcal{K}\subset\mathbb{R}^d$} compact, and a continuous {$\mathbb{R}^d$}-valued function $V$, with
{$\lim_{\Vert{x}\Vert\to\infty} V(x) = \infty$}, such that under any non-anticipative {$\{\theta(n)\}$},
\begin{enumerate}[(i)]
\item 
{$\sup_n \mathbb{E}[V(Y_n)^2] < \infty$} and
\item 
{$\mathbb{E}[V(Y_{n+1})|\mathcal{F}_n] \leqslant V(Y_n) - \epsilon_0$}, when {$Y_n\notin\mathcal{K}$}, {$n\geqslant0$}.
\end{enumerate} 
\end{assumption}

Assumption~\ref{differentiable} is a technical requirement, whereas Assumption~\ref{lyapunov} is used to show the stability of the scheme.
Assumption~\ref{lyapunov} will not be required, for instance, if the single-stage cost function $h$ is bounded in addition. 

\subsection{Smoothed Functionals}
Given any function {$f:C\mapsto\mathbb{R}$}, its smoothed functional is defined as
\begin{equation}
\label{SF1}
S_{\beta}[f(\theta)] = \int\limits_{-\infty}^{\infty}G_{\beta}(\eta)f(\theta-\eta)\mathrm{d}\eta
 = \int\limits_{-\infty}^{\infty}G_{\beta}(\theta-\eta)f(\eta)\mathrm{d}\eta,
\end{equation}
where {$G_{\beta}:\mathbb{R}^N\mapsto\mathbb{R}$} is a kernel function. 

The idea behind using smoothed functionals is that if {$f(\theta)$} is not well-behaved, \textit{i.e.}, it 
has a fluctuating character, then {$S_{\beta}[f(\theta)]$} has less fluctuations for
appropriate values of $\beta$. This ensures that
any optimization algorithm with objective function {$f(\theta)$} does not get stuck at any local minimum, but 
converges to the global minimum. The parameter {$\beta$} controls the degree of smoothness.
Rubinstein~\cite{rubinstein} has shown that the SF algorithm achieves these properties if the kernel 
function satisfies the following sufficient conditions:
\begin{enumerate}[(P1)]
\item {$G_{\beta}(\eta) = \frac{1}{{\beta}^N}G(\frac{\eta}{\beta})$}, \\where {$G(\frac{\eta}{\beta}) = G_1(\frac{\eta}{\beta})
= G_1({\frac{\eta^{(1)}}{\beta}},{\frac{\eta^{(2)}}{\beta}},\ldots,{\frac{\eta^{(N)}}{\beta}})$}.
\item {$G_{\beta}(\eta)$} is piecewise differentiable in {$\eta$}.
\item {$G_{\beta}(\eta)$} is a probability distribution function, \\\textit{i.e.}, {$S_{\beta}[f(\theta)] = \mathbb{E}_{G_{\beta}(\eta)}[f(\theta-\eta)]$}.
\item {$\lim_{\beta\to0}G_{\beta}(\eta) = \delta(\eta)$}, the Dirac delta function.
\item {$\lim_{\beta\to0}S_{\beta}[f(\theta)] = f(\theta)$}.
\end{enumerate}

The normal distribution satisfies the above conditions, and has been used as a kernel by Katkovnik~\cite{katkovnik}. 

Based on \eqref{SF1}, a form of gradient estimator has been derived in \cite{bhatnagar3} which is given by
\begin{equation}
\label{G_estimate1}
\nabla_{\theta}[J(\theta)]\approx\frac{1}{\beta ML}\sum_{n=0}^{M-1}\sum_{m=0}^{L-1}{\eta(n)h(Y_m)}
\end{equation}
for large $M$, $L$ and small {$\beta$}. The process {$\{Y_m\}$} is governed by parameter {$(\theta(n)+\beta\eta(n))$}, 
where {$\theta(n)\in C\subset\mathbb{R}^N$} is obtained through an iterative scheme. {$\eta(n)$} is a $N$-dimensional 
vector composed of i.i.d. {$\mathcal{N}(0,1)$}-distributed random variables.

%====================Section: q-Gaussian for Smoothed Functionals============
\section{\lowercase{$q$}-Gaussian for Smoothed Functionals}
\label{validity}
\begin{proposition}
The $q$-Gaussian distribution satisfies the kernel properties \textnormal{(P1) -- (P5)} for all {$q<3$}, {$q\neq1$}. 
\end{proposition}
\begin{proof}
\begin{enumerate}[(P1)]
\item 
From~\eqref{Gq:formula}, it is evident that
{$G_{q,\beta}(\eta) = \displaystyle\frac{1}{{\beta}^N}G_q\left(\frac{\eta}{\beta}\right)$}.
\item 
For {$1<q<3$}, {$\left(1-{\frac{(1-q)}{(3-q)\beta^2}{\Vert}{\eta}{\Vert}^2}\right)>0$},
for all {$\eta\in\mathbb{R}^N$}. 
\begin{align}
&\text{Hence, } G_{q,\beta}(\eta) = {\frac{1}{\beta^NK_{q,N}}}\left(1-{\frac{(1-q)}{(3-q)\beta^2}{\Vert}
\eta{\Vert}^2}\right)^{\frac{1}{1-q}}.	\nonumber
\\&\text{Thus, }\nabla_{\eta}G_{q,\beta}(\eta) = -{\frac{2\eta}{(3-q)\beta^2}} \frac{G_{q,\beta}(\eta)}
{\left(1-{\frac{(1-q)}{(3-q)\beta^2}{\Vert}\eta{\Vert}^2}\right)}.
\label{grad}
\end{align}
For {$q<1$}, when {$\Vert\eta\Vert^2<\frac{(3-q)\beta^2}{(1-q)}$} , we have
\begin{displaymath}
\left(1-{\frac{(1-q)}{(3-q)\beta^2}{\Vert}{\eta}{\Vert}^2}\right)>0. 
\end{displaymath}
So,~\eqref{grad} holds.
On the other hand, when {$\Vert\eta\Vert^2\geqslant\frac{(3-q)\beta^2}{(1-q)}$}, we have 
{$\left(1-{\displaystyle\frac{(1-q)}{(3-q)\beta^2}{\Vert}{\eta}{\Vert}^2}\right)\leqslant0$},
which implies {$G_{q,\beta}(\eta) = 0$} and, {$\nabla_{\eta}G_{q,\beta}(\eta) = 0$}.
\\Thus, {$G_{q,\beta}(\eta)$} is differentiable for {$q>1$}, and piecewise differentiable for {$q<1$}.
\item
{$G_{q,\beta}(\eta)$} is a distribution for {$q<3$} and hence, the corresponding SF {$S_{q,\beta}(.)$},
which is parameterized by both $q$ and $\beta$ can be written as
\begin{displaymath}
S_{q,\beta}[f(\theta)] = \mathbb{E}_{G_{q,\beta}(\eta)}[f(\theta-\eta)].
\end{displaymath}
\item
As {$\beta\to0$},  {$G_{q,\beta}(0) = \displaystyle\frac{1}{\beta^NK_{q,N}}\to\infty$}.
But, we have {$\displaystyle\int\limits_{\mathbb{R}^N} G_{q,\beta}(\eta)\mathrm{d}\eta = 1$} for {$q<3$}.
So, {$\displaystyle\lim_{\beta\to0}G_{q,\beta}(\eta) = \delta(\eta)$}.
\item
It follows from dominated convergence theorem that
\begin{align*}
\lim_{\beta\to0}S_{q,\beta}[f(\theta)] &=\int\limits_{-\infty}^{\infty}{\lim_{\beta\to0}G_{q,\beta}(\eta)f(\theta-\eta)d\eta}
\\&=\int\limits_{-\infty}^{\infty}{\delta(\eta)f(\theta-\eta)\mathrm{d}\eta}
=f(\theta). 
\end{align*}
\end{enumerate}
\vspace{-1cm} \[\qedhere\]
\end{proof}

Our objective is to estimate {$\nabla_{\theta}J(\theta)$} using the SF approach. The existence of 
{$\nabla_{\theta}J(\theta)$} is due to Assumption~\ref{differentiable}. Now, 
\begin{displaymath}
\nabla_{\theta}J(\theta) = \left[\nabla_{\theta}^{(1)}J(\theta) \quad\nabla_{\theta}^{(2)}J(\theta)
 \quad\ldots \quad\nabla_{\theta}^{(N)}J(\theta)\right]^T.
\end{displaymath}
\\Let us define,
{$\Omega_q = \left\{\eta\in\mathbb{R}^N:\Vert\eta\Vert^2 < \frac{(3-q)\beta^2}{(1-q)}\right\}$} for 
{$q<1$}, and {$\Omega_q = \mathbb{R}^N$} for {$1<q<3$}. It is evident that {$\Omega_q$} is the support set 
for the $q$-Gaussian distribution with $q$-variance {$\beta^2$}.
\\Define the SF for gradient of average cost as 
\begin{align*}
D_{q,\beta}[J(\theta)] &= \Big[S_{q,\beta}[\nabla_{\theta}^{(1)}J(\theta)]
\quad\ldots\quad S_{q,\beta}[\nabla_{\theta}^{(N)}J(\theta)] \Big]^T
\\&=\int\limits_{\mathbb{R}^N}G_{q,\beta}(\theta-\eta){\nabla_{\eta}J(\eta)}\mathrm{d}\eta \;.
\end{align*}
It follows from integration by parts and the definition of {$\Omega_q$},
\begin{align*}
D_{q,\beta}[J(\theta)]&=\int\limits_{\Omega_q}\nabla_{\eta}G_{q,\beta}(\eta)J(\theta-\eta)\mathrm{d}\eta \;.
\end{align*}
Substituting {$\bar{\eta}=-\frac{\eta}{\beta}$}, we have
\begin{align}
D_{q,\beta}[J(\theta)]
&=\int\limits_{\Omega_q}{\frac{2}{(3-q)\beta}}\frac{\bar{\eta}J(\theta+\beta\bar{\eta})}{\left(1-\frac{(1-q)}{(3-q)}
\Vert\bar{\eta}\Vert^2\right)}G_q(\bar{\eta})\mathrm{d}\bar{\eta}	\nonumber
\\&=\frac{2}{\beta(3-q)}\mathbb{E}_{G_q(\bar{\eta})}\left[\frac
{\bar{\eta}J(\theta+\beta\bar{\eta})}{\left(1-\frac{(1-q)}{(3-q)}\Vert\bar{\eta}\Vert^2\right)}\right]\enspace.
\label{D:formula} 
\end{align}

We first state the following lemma which will be required to prove the result in Proposition~\ref{D:convergence}.
\begin{lemma}
\label{q-expect}
Let {$f:\mathbb{R}^N\mapsto\mathbb{R}$} be a function defined over a standard $q$-Gaussian 
distributed random variable {$X\in\mathbb{R}^N$},
\begin{displaymath}
\textit{i.e.,} \langle{X}\rangle_q = 0
\text{ and } \langle{XX^T}\rangle_q = I_{N{\times}N},
\end{displaymath}
\begin{displaymath}
\text{then, }\quad\langle{f(X)}\rangle_q = \frac{1}{\Lambda_q}\mathbb{E}_{G_q(X)} \left[
\frac{f(X)}{1-\frac{(1-q)}{(3-q)}\Vert{X}\Vert^2}\right]\enspace,\quad\qquad
\end{displaymath}
where {$\Lambda_q = \left[(K_{q,N})^{q-1}\displaystyle\int_{\mathbb{R}^N}[G_q(x)]^q\mathrm{d}x\right]$},
{$K_{q,N}$} being the normalizing constant for N-variate $q$-Gaussian.
\end{lemma}

\begin{proof}
From ~\eqref{q-expect-defn}
\begin{align*}
\langle{f(X)}\rangle_q
&= \frac{\displaystyle\int_{\mathbb{R}^N}f(x)[G_{q}(x)]^q\mathrm{d}x}
{\displaystyle\int_{\mathbb{R}^N}[G_{q}(x)]^q\mathrm{d}x}
\\&= \frac{1}{\Lambda_q K_{q,N}}\int\limits_{\mathbb{R}^N}{f(X)}
\left(1-\frac{(1-q)\Vert{x}\Vert^2}{(3-q)}\right)_+^{\frac{q}{1-q}}\mathrm{d}x	\quad
\\&= \frac{1}{\Lambda_q}\int\limits_{\Omega_q}\frac{f(x)}{\left(1-\frac{(1-q)}{(3-q)}
\Vert{x}\Vert^2\right)}G_q(x)\mathrm{d}x
\\&= \frac{1}{\Lambda_q}\mathbb{E}_{G_q(X)} \left[\displaystyle\frac{f(X)}{1-\frac{(1-q)}{(3-q)}
\Vert{X}\Vert^2}\right]\enspace.
\end{align*}
\vspace{-1.5cm} \[\qedhere\]
\end{proof}

\begin{proposition}
\label{D:convergence}
For a given {$q<3$}, {$q\neq1$}, as {$\beta\to0$}, SF for the gradient converges 
to a scaled version of the gradient, 
\begin{displaymath}
\text{i.e., }\left\Vert D_{q,\beta}[J(\theta)]-\frac{2\Lambda_q}{(3-q)}\nabla_{\theta}J(\theta) \right
\Vert \to 0 \text{ as  }\beta \to 0.
\end{displaymath}
\end{proposition}

\begin{proof}
For small {$\beta$}, using Taylor series expansion,
\begin{displaymath}
 J(\theta+\beta\bar{\eta}) = J(\theta) + \beta\bar{\eta}^T{\nabla_{\theta}J(\theta)} +
\frac{1}{2}\beta^2\bar{\eta}^T{\nabla_{\theta}^2J(\theta)}\bar{\eta} + o(\beta^2) 
\end{displaymath}
By Lemma~\ref{q-expect}, 
\begin{align*}
&D_{q,\beta}[J(\theta)] 
= \frac{2\Lambda_q}{\beta(3-q)}\bigg\langle\bar{\eta}J(\theta+\beta\bar{\eta})\bigg\rangle_q
\\&= \frac{2\Lambda_q}{\beta(3-q)}\bigg[\Big\langle{\bar{\eta}}\Big\rangle_q{J(\theta)} +
\beta\Big\langle{\bar{\eta}\bar{\eta}^T}\Big\rangle_q{\nabla_{\theta}J(\theta)} + 
\\&\qquad\qquad\qquad\qquad\qquad\frac{1}{2}\beta^2\Big\langle{\bar{\eta}\bar{\eta}^T{\nabla_{\theta}^2J(\theta)}\bar{\eta}}\Big\rangle_q + o(\beta^2)\bigg]
\\&= \frac{2\Lambda_q}{(3-q)}\left[{\nabla_{\theta}J(\theta)} +
\beta\left(\frac{1}{2}\Big\langle{\bar{\eta}\bar{\eta}^T{\nabla_{\theta}^2J(\theta)}\bar{\eta}}\Big\rangle_q + o(\beta)\right)\right]
\end{align*}
Thus, {$D_{q,\beta}[J(\theta)]\to\left(\displaystyle\frac{2\Lambda_q}{(3-q)}{\nabla_{\theta}J(\theta)}\right)$} as {$\beta\to0$}.
\end{proof}

As a consequence of the Proposition \ref{D:convergence}, for large $M$ and small {$\beta$}, 
the form of gradient estimate suggested by~\eqref{D:formula} is
\begin{equation}
\nabla_{\theta}[J(\theta)]\approx\frac{1}{\Lambda_q\beta M}\sum_{n=0}^{M-1}\left[\frac{\bar{\eta}(n)
J(\theta(n)+\beta\bar{\eta}(n))}{\left(1-\frac{(1-q)}{(3-q)}\Vert\bar{\eta}(n)\Vert^2\right)}\right].
\end{equation}
Using an approximation of~\eqref{Jdefn}, for large $L$, we can write the above equation as
\begin{equation}
\label{estimate1}
\nabla_{\theta}[J(\theta)]\approx\frac{1}{\Lambda_q\beta ML}\sum_{n=0}^{M-1}\sum_{m=0}^{L-1}
\frac{\bar{\eta}(n)h(Y_m)}{\left(1-\frac{(1-q)}{(3-q)}\Vert\bar{\eta}(n)\Vert^2\right)},
\end{equation}
where {$\{Y_m\}$} is governed by parameter {$(\theta(n)+\beta\bar{\eta}(n))$}.

However, since {$\Lambda_q>0$}, {$\Lambda_q$} need not be explicitly determined  as estimating 
{$[\Lambda_q\nabla_{\theta}J(\theta)]$} instead of {$\nabla_{\theta}J(\theta)$} does not affect the gradient descent approach.
As a special case, for {$q=1$}, we have {$\Lambda_q=1$} from definition. %Thus, substituting {$q=1$} in \eqref{estimate1},
Hence, we obtain the same form as in \eqref{G_estimate1}.

%====================Section: Proposed Algorithms============
\section{Proposed Algorithms}
\label{algorithm}
In this section, we propose a two-timescale algorithm corresponding to the estimate obtained in~\eqref{estimate1}.

The $q$-Gaussian distributed parameters ({$\eta$}) have been generated in the algorithm using the method
proposed in \cite{thistleton}.
\\Let \{{$a(n)$}\}, \{{$b(n)$}\} be two step-size sequences satisfying
\begin{assumption}
\label{stepsize}
{$a(n) = o(b(n))$}, {$\displaystyle\sum_{n=0}^{\infty}a(n) = \sum_{n=0}^{\infty}b(n) = \infty$}, 
and {$\displaystyle\sum_{n=0}^{\infty}a(n)^2, \sum_{n=0}^{\infty}b(n)^2 < \infty$}.
\end{assumption}
For {$\theta = (\theta^{(1)},\ldots,\theta^{(N)})^T \in\mathbb{R}^N$}, let {$\Gamma(\theta) = \big(\Gamma(\theta^{(1)}),\ldots,$}
 {$\Gamma(\theta^{(N)})\big)^T$}
represent the projection of {$\theta$} onto the set $C$. \{{$Z^{(i)}(n), i=1,\ldots,N$}\}{$_{n\in\mathbb{N}}$} are  quantities used to estimate 
{$[\Lambda_q\nabla_{\theta}J(\theta)]$} via the recursions below.

\begin{algorithm}
\caption{} 
\begin{algorithmic}[1]
\STATE Fix $M$, $L$, $q$ and $\beta$.
\STATE Set {$Z^{(i)}(0) = 0, i=1,\ldots,N$}.
\STATE Fix parameter vector {$\theta(0)=(\theta^{(1)}(0),\ldots,\theta^{(N)}(0))^T$}.
\FOR {{$n=0$} to {$M-1$}}
\STATE Generate i.i.d. standard $q$-Gaussian distributed random variables 
{$\eta^{(1)}(n),\ldots,\eta^{(N)}(n)$} and set \\{$\eta(n)=(\eta^{(1)}(n),\ldots,\eta^{(N)}(n))^T$}.
\FOR {{$m=0$} to {$L-1$}}
\STATE Generate the simulation {$Y_{nL+m}$} governed with parameter 
{$(\theta(n)+\beta\eta(n))$}.
\FOR {{$i=1$} to {$N$}}
\STATE {$Z^{(i)}(nL+m+1) = (1-b(n))Z^{(i)}(nL+m)$}
\\{$\qquad\qquad\qquad\qquad + b(n)\left[\frac{\eta^{(i)}(n)h(Y_{nL+m})}{\beta\left(1-\frac{(1-q)}{(3-q)}\Vert{\eta}(n)\Vert^2\right)}\right]$}. 
\ENDFOR
\ENDFOR
\FOR {{$i=1$} to {$N$}}
\STATE
{$\theta^{(i)}(n+1) = \Gamma\left(\theta^{(i)}(n) - a(n)Z^{(i)}(nL)\right)$}.
\ENDFOR
\STATE Set {$\theta(n+1)=(\theta^{(1)}(n+1),\ldots,\theta^{(N)}(n+1))^T$}.
\ENDFOR
\STATE Output {$\theta(M)$} as the final parameter vector.
\end{algorithmic}
\end{algorithm}

%====================Section: Numerical Experiments============
\section{Numerical Experiment}
\label{results}
\subsection{Numerical Setting}
We consider a two-node network of {$M/G/1$} queues with feedback. The setting here is somewhat similar to that considered in \cite{bhatnagar3}.
Nodes 1 and 2 are fed with independent Poisson external arrival processes with rates {$\lambda_1=0.2$} and {$\lambda_2=0.1$}, respectively.
After departing from Node-1, customers enter Node-2. Once the service at Node-2 is completed, a customer either leaves the system with probability
{$p=0.4$} or joins Node-1. The service time processes of the two nodes, {$\{S_n^1(\theta_1)\}_{n\geqslant1}$} and 
{$\{S_n^2(\theta_2)\}_{n\geqslant1}$}, respectively, are defined as
\begin{equation}
S_n^i(\theta_i) = U_i(n)\frac{\left(1+\Vert{\theta_i(n) - \bar{\theta}_i}\Vert^2\right)}{R_i} \quad i=1,2, n\geqslant1,
\end{equation}
where {$R_1=10$} and {$R_2=20$} are constants. Here, {$U_1(n)$} and {$U_2(n)$} are independent samples drawn from uniform distribution
on (0,1). Service time of each node depends on the {$N_i$}-dimensional tunable parameter vector {$\theta_i$}, whose individual components
lie in a certain interval {$[(\theta_i^{(j)})_{min},(\theta_i^{(j)})_{max}]$}, {$j=1,\ldots,N_i$}, {$i=1,2$}. 
{$\theta_i(n)$} represents the $n^{th}$ update of parameter vector at Node-$i$, and {$\bar{\theta}_i$}
represents the target vector. 

The cost function is chosen to be the sum of the two queue lengths at any instant. For the cost to be minimum, {$S_n^i(\theta_i)$} should
be minimum, and hence, we should have {$\theta_i(n)=\bar{\theta}_i$}, {$i=1,2$}. We denote 
{$\theta = (\theta_1^{(1)},..,\theta_1^{(N_1)},\theta_2^{(1)},..,\theta_2^{(N_2)})\in\mathbb{R}^N$},
and {$\bar{\theta} = (\bar\theta_1^{(1)},..,\bar\theta_1^{(N_1)},\bar\theta_2,..,\bar\theta_2^{(N_2)})\in\mathbb{R}^N$}, where $N$=$N_1$+$N_2$.
For the simulations, we use the following values of parameters:
\\(1) {$N_1=N_2=2$},
\\(2) {$(\theta_i^{(j)})_{min} = 0$}, {$(\theta_i^{(j)})_{max} = 5$} for all {$i,j$}, \textit{i.e.}, {$C = [0,5]^N$}.
\\(3) {$\theta^{(j)}(0) = 5$}, {$\bar{\theta}^{(j)} = 1$} for {$j=1,2,\ldots,N$},
\\(4) {$M=10000$}, {$L=100$},
\\(5) {$a(n) = 1/n$}, {$b(n) = 1/n^{2/3}$}.

\subsection{Simulation Results}
Simulations are performed by varying the parameters $q$ and $\beta$. 
We compare the performance of our algorithm with the SF algorithm proposed in 
\cite{bhatnagar3}, which uses Gaussian smoothing. 
The Euclidian distance between {$\theta(n)$} and {$\bar{\theta}$} is chosen as 
the performance measure as this gives the proximity of the updates to the global optimum.
For each case, the results are averaged over $20$ independent trials. 
Figure~1 shows that with same {$\beta$}, $q$-SF converges faster than SF algorithm for some $q$'s.
Table~I presents a detailed comparison for different values of $q$ and $\beta$. 

\begin{figure}[h]
\centering
\includegraphics[height=3.6cm]{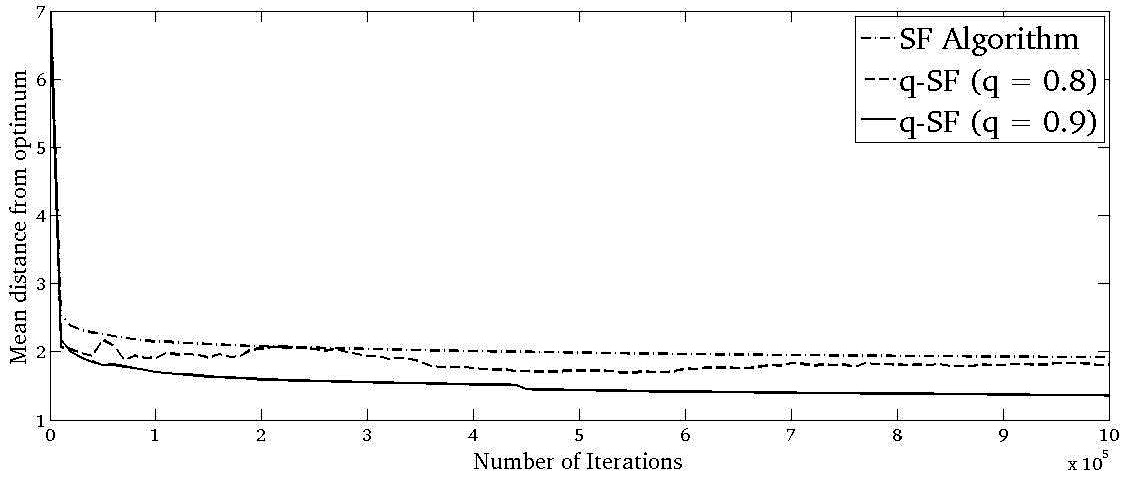}
\caption{Convergence behavior of the algorithm for {$\beta=0.25$}.}
\end{figure}
\vspace{-5mm}
\begin{table}[ht]
\footnotesize
\label{GqSF1_error}
\begin{tabular}{|c||c|c|c|c|c|c|c|c|c|c|c|c|c|c|c|}
\hline
{$q$}{$\diagdown$}{$\beta$} 	&	0.0005	&	0.005	&	0.05	&	0.1	&	0.25		&	0.5		&	1		&	2.5		\\
\hline
\hline
0	&	3.62		&	2.78		&	2.86		&	3.10	&	3.08		&	\textbf{2.82}	&	3.51		&	3.20		\\
0.5	&	4.05		&	2.70		&	2.68		&	2.90	&	2.91		&	3.15		&	2.95		&	3.20		\\
0.6	&	3.82		&	2.91		&	2.83		&	3.16	&	3.03		&	\textbf{\underline{2.78}}	&	2.90		&	3.53		\\
0.7	&	4.37		&	2.75		&	2.57		&	2.60	&	\textbf{\underline{2.19}}	&	2.97		&	2.93		&	2.93		\\
0.8	&	3.97		&	2.47		&	2.98		&	2.42	&	\textbf{2.48}	&	\textbf{2.91}	&	2.72		&	2.90		\\
0.9	&	2.66		&	2.06		&	\textbf{\underline{2.14}}	&	2.48	&	\textbf{2.43}	&	\textbf{\underline{2.78}}	&	\textbf{\underline{2.07}}	&	\textbf{\underline{1.18}}	\\
1.1	&	2.19		&	\textbf{1.81}	&	\textbf{2.19}	&	2.93	&	2.78		&	3.07		&	2.78		&	1.59		\\
1.2	&	\textbf{1.85}	&	\textbf{1.81}	&	\textbf{2.21}	&	2.75	&	3.27		&	3.31		&	3.28		&	1.78		\\
1.3	&	2.32		&	\textbf{1.77}	&	2.69		&	3.18	&	3.55		&	3.77		&	3.46		&	2.10		\\
1.4	&	\textbf{1.69}	&	\textbf{\underline{1.67}}	&	\textbf{2.42}	&	2.98	&	3.46		&	3.96		&	3.92		&	2.45		\\
1.5	&	2.34		&	2.02		&	2.89		&	2.94	&	3.88		&	4.00		&	3.75		&	2.51		\\
1.6	&	\textbf{1.80}	&	\textbf{1.76}	&	3.15		&	3.23	&	4.09		&	3.90		&	3.74		&	2.95		\\
2	&	\textbf{\underline{1.65}}	&	2.10		&	3.47		&	4.46	&	4.64		&	5.10		&	4.60		&	4.23		\\
2.5	&	\textbf{1.97}	&	2.65		&	3.98		&	4.66	&	5.77		&	6.01		&	6.14		&	5.74		\\
\hline
\hline
SF	&	2.09		&	1.85		&	2.52		&	\underline{2.09}	&	2.77		&	2.96		&	2.65		&	1.31		\\
\hline
\end{tabular}
\caption{Performance (mean distance from optimum).} 
\end{table}

The cases where $q$-SF outperforms SF are highlighted, and for each $\beta$, 
the best result is underlined. It can be observed that
for smaller {$\beta$}, $q$-SF with $q>1$ performs better than SF, but for
larger {$\beta$}, better performance can be obtained with {$q<1$}. 
So, as {$\beta$} increases, smaller $q$'s prove to be better.
As per observations, $q=0.9$ performs better than Gaussian in 63\% cases,
and also gives the least distance in most of the cases (50\%).

The results show that there are some values of {$q\neq1$} for which
we can reach closer proximity of the global minimum with the proposed algorithm
than the SF case. This can be contributed to the power-law tail
of $q$-Gaussian which allows better control over the level of smoothing.
There is an additional improvement provided by {$\Lambda_q$}, which can be expressed as
\begin{equation}
\Lambda_q = \mathbb{E}_{G_q(X)} \left[\left(1-\frac{(1-q)}{(3-q)}\Vert{X}\Vert^2\right)^{-1}\right]\enspace.
\end{equation}
For {$q>1$}, the term inside bracket is always less than 1, which implies {$\Lambda_q<1$}, 
whereas {$\Lambda_q>1$} for {$q<1$}. Thus the gradient descent is faster for {$q<1$}, which
leads to faster convergence.

We also note that for high $q$, the algorithm does not converge for larger $\beta$.
So we may claim that the region of stability of $q$-SF, given by $\beta_0$ (see Theorem~\ref{thm}), decreases as $q$ increases.

%====================Section: Convergence Analysis============
\section{Sketch of Convergence Analysis}
\label{convergence}
Here, we give a sketch of the proof of convergence of the proposed algorithm. 
We just state the important results. The proofs will be given in a longer version of the paper. 

Let {$\mathcal{F}(l) = \sigma\big(\tilde{\theta}^{(i)}(k), \tilde{\eta}^{(i)}(k), Y_k, k\geqslant l, i = 1,\ldots, N\big)$}, {$l\geqslant1$}
denote the {$\sigma$}-fields generated by the above mentioned quantities, where {$\tilde{\theta}^{(i)}(k) = \theta^{(i)}(n)$} and
{$\tilde{\eta}^{(i)}(k) = \eta^{(i)}(n)$} for {$i = 1,\ldots N$}, {$nL\leqslant k <(n+1)L$}. Define {$\{\tilde{b}(n)\}_{n\geqslant0}$}
such that {$\tilde{b}(n) = b(\left[\frac{n}{L}\right])$}, where {$[x]$} is the integer part of $x$. Thus,
{$\displaystyle\sum_{n=0}^{\infty}\tilde{b}(n) = \infty$}, 
{$\displaystyle\sum_{n=0}^{\infty}\tilde{b}(n)^2 < \infty$} and {$\tilde{b}(n) = o(b(n))$}.

With the above notation, substituting {$p = nL+m$} we can rewrite Step 9 of our algorithm in terms 
of {$\tilde{b}(p)$}, {$\tilde{\theta}^{(i)}(p)$} and {$\tilde{\eta}^{(i)}(p)$}. We 
define the sequences {$\{M^{(i)}(p)\}_{p\geqslant1}$}, {$i = 1,\ldots N$},
\begin{align}
M^{(i)}(p) &= \sum_{k=1}^p \tilde{b}(k) \left(\frac{\tilde\eta^{(i)}(k)h(Y_k)}{\beta\left(1-\frac{(1-q)}{(3-q)}\Vert{\tilde{\eta}}(n)\Vert^2\right)} 
\right. \nonumber	\\ & -\left.
\mathbb{E}_{G_q}\left[\frac{\tilde\eta^{(i)}(k)h(Y_k)}{\beta\left(1-\frac{(1-q)}{(3-q)}\Vert{\tilde\eta}(k)\Vert^2\right)}\Bigg|\mathcal{F}(k-1)\right]\right)
\end{align}

\begin{lemma}
The sequences {$\{M^{(i)}(p), \mathcal{F}(p)\}_{p\geqslant1}$}, {$i = 1,$} {$2,\ldots N$} are almost surely convergent martingale sequences. 
\end{lemma}

Consider the following ordinary differential equations:
\begin{align}
\dot{\theta}(t) &= 0,
\\ \dot{Z}(t) &= \frac{(3-q)}{2}D_{q,\beta}[J(\theta)] - Z(t).
\end{align}
\begin{lemma}
\label{lem}
The sequence of updates {$\{Z(p)\}$} is uniformly bounded with probability 1.
\end{lemma}
\begin{lemma}
\label{Z2D}
For a given {$q<3$}, {$q\neq1$}, with probability 1
{$\left\Vert Z(nL) - \frac{(3-q)}{2}D_{q,\beta}[J(\theta(n))] \right\Vert \to 0$} as {$n\to\infty$}.
\end{lemma}

The following corollary follows directly from Proposition~\ref{D:convergence} and Lemma~\ref{Z2D} by triangle inequality.
\begin{corollary}
Given a particular {$q<3$}, with probability 1, as {$n\to\infty$} and {$\beta\to 0$},
{$\left\Vert Z(nL)-\Lambda_q\nabla_{\theta}J(\theta) \right\Vert \to 0$} 
\end{corollary}

Now, finally considering the ODE corresponding to the slowest timescale recursion:
\begin{equation}
\label{ode}
\dot{\theta}(t) = \tilde{\Gamma}\big(-\Lambda_q\nabla_{\theta}J(\theta(t))\big),
\end{equation}
where {$\tilde{\Gamma}(f(x))=\lim_{\epsilon\to0}\left(\frac{\Gamma(x + \epsilon f(x)) - x}{\epsilon}\right)$} 
for any bounded, continuous function {$f:\mathbb{R}^N\to\mathbb{R}^N$}. The stable points of \eqref{ode} lie in 
the set {$S = \big\{\theta\in C : \tilde{\Gamma}\big(-\Lambda_q\nabla_{\theta}J(\theta(t))\big) = 0\big\}$}.
Given {$\delta>0$}, we define {$S^{\delta}= \big\{ \theta\in C : \Vert{\theta-\theta_0}\Vert < \delta, \theta_0\in S\big\}$}.
\begin{theorem}
\label{thm}
Under Assumptions~\ref{differentiable}~--~\ref{stepsize}, given {$q<3$}, {$q\neq1$} and {$\delta>0$}, {$\exists\beta_0 >0$} such that
for all {$\beta\in(0,\beta_0]$}, the sequence {$\{\theta(n)\}$} obtained using the $q$-SF algorithm converges to a point in {$S^\delta$}
with probability $1$ as {$n\to\infty$}.
\end{theorem}

%====================Section: Conclusion============
\section{Conclusion}
\label{conclusion}
The $q$-Gaussian exhibits power-law behavior, which gives a better control over smoothing of functions as compared to normal distribution. 
We have 
extended the Gaussian smoothed functional gradient estimation approach to $q$-Gaussians, and developed an optimization algorithm
based on this. We have also presented results illustrating that for some values of $q$, our algorithm
performs better than the SF algorithm~\cite{bhatnagar3}.

%====================Section: References============

\end{document}